\documentclass[amssymb,prb,onecolumn,floats,amsmath]{revtex4}
\usepackage{bm}
\usepackage{graphicx}

\begin{document}
%\date{\today}
%################################################################# TITLE

\title{Designing quantum dots for solotronics}

%################################################################# AUTHORS

\author{J.~Kobak}
\author{T.~Smole\'nski}
\author{M.~Goryca}
\author{M.~Papaj}
\author{K.~Gietka}
\author{A.~Bogucki}
\author{M.~Koperski}
\author{J.-G.~Rousset}
\author{J.~Suf\mbox{}fczy\'{n}ski}
\author{E.~Janik}
\author{M.~Nawrocki}
\author{A.~Golnik}
\author{P.~Kossacki}
\author{W.~Pacuski}
\email{Wojciech.Pacuski@fuw.edu.pl}

\affiliation{Institute of Experimental Physics, Faculty of Physics, University of Warsaw, Ho\.za 69, 00-681 Warsaw, Poland}

%####################################################### ABSTRACT

\maketitle

%\begin{abstract}
\textbf{Solotronics, optoelectronics based on solitary dopants, is an emerging field of research and technology reaching the ultimate limit of miniaturization. It aims at exploiting quantum properties of individual ions or defects embedded in a semiconductor matrix. As already shown, optical control of a spin of a magnetic ion is feasible employing photo-generated carriers confined in a quantum dot. A non-radiative recombination, regarded as a severe problem, limited development of quantum dots with magnetic ions. Our photoluminescence studies on, so far unexplored, individual CdTe dots with single cobalt ions and individual CdSe dots with single manganese ions show, however, that even if energetically allowed, the single ion related non-radiative recombination is negligible in such zero-dimensional structures. This opens solotronics for a wide range of even not yet considered systems. Basing on the results of our single spin relaxation experiments and on the material trends, we identify optimal magnetic ion-quantum dot systems for implementation of a single-ion based spin memory.}\\
%\end{abstract}

%##################################################### INTRODUCTION

%\section* {Introduction}

The term solotronics\cite{Flatte_NM11} has been introduced to describe recent advances\cite{Awschalom_S13} in fabricating and operating semiconductor optoelectronic devices based on single dopants or defects, for applications in computer memories, quantum computation and  on demand photon sources. The most advanced solotronics technology has been developed for nitrogen-vacancy (N-V) defect centers in diamond, for which it has been shown that quantum states can be prepared and readout, and spin can be manipulated using microwave and optical transitions\cite{Jelezko_PRL04,Balasubramanian_N09,Dolde_NP13,Yale2013}.%Gaebel_N06,Neumann_NPh10,Fuchs_PRL12,Redman_PRL91,

Defects similar to N-V centers were observed also in semiconductors\cite{Weber_PNAS10} such as SiC\cite{Koehl_N11}. The SiC is more compatible than diamond to the present semiconductor-based technology, however, due to the weak coupling of free carriers to defect centers, it does not allow for electrically controlled operation. More promising in this view is another solotronic system: a single magnetic ion embedded in a semiconductor quantum dot (QD)\cite{Besombes_PRL04, Kudelski_PRL07,Goryca_PRL09QDs}. Here, the spin state of the single ion can be prepared and manipulated both electrically\cite{Leger_PRL06,Fernandez_PRL07} and optically\cite{Gall_PRL09, Goryca_PRL09QDs,Baudin_PRL11} through injection of spin polarized carriers. The $s$,$p$-$d$ exchange coupling between the magnetic ion and the band carrier enables an unambiguous readout of the spin projection of the ion from the energy and polarization of a photon emitted from the QD\cite{Besombes_PRL04,Kudelski_PRL07}. The ease of optical addressing of individual QDs enables operation on the level of single ions\cite{Gall_PRL09, Goryca_PRL09QDs}. Multiple magnetic ions can be coupled by carriers in one QD\cite{Besombes_PRB12,Beaulac_S09,Klopotowski_PRB11} or by QDs coupling through tunneling carriers\cite{Goryca_PRL09QDs} or photonic structures\cite{Andrade_PRB12}. The use of  semiconductor heterostructures opens a huge space for testing new ideas for single-ion spin operation, as it offers a band gap and strain engineering, tuning energies of optical and microwave transitions, Fermi level manipulation, and integration to p-i-n structures.

A severe limitation of QDs doped with transition metal ions was attributed to the efficient recombination channel introduced by magnetic ions\cite{Oka_JL99,Tang_JCG03,Lee_PRB05,Beaulac_NL08,Beaulac_NL08XPL,Zhong_Sp09,Bussian_NM09,Yamamoto_JAP10,Pandey_NM12}, when the exciton energy is higher than the intra-ionic transition energy, which should result in quenching of exciton emission. Therefore, the only QD systems with single magnetic ions considered so far were those where intra-ionic transitions energies  exceed exciton energy, namely Mn$^{2+}$ embedded in CdTe/ZnTe and InAs/GaAs QDs.\cite{Besombes_PRL04,Leger_PRL06,Gall_PRL09,Goryca_PRB10,Goryca_PRL09QDs,Fernandez_PRL07, Kudelski_PRL07,Krebs_PRB09, Baudin_PRL11, Gall_PRB12}
On the other hand, incorporation of magnetic ions such as Cr, Fe, Co, Ni or Cu would bring physical properties like orbital momentum, reduced number of spin states, sensitivity to a local strain, Jahn-Teller effect, or isotopes with zero nuclear spin, offering additional degrees of freedom for designing quantum states. Extending the studies of single magnetic ions to other QD systems such as CdSe, ZnSe, CdS, ZnS, ZnO, GaN or other wide gap semiconductors would offer, in turn, increased PL efficiency at higher temperatures, enhancement of the exchange interaction within excitons and between excitons and ions, or reduction of spin-orbit coupling and resulting spin relaxation rates.

Here, for the first time, we report on  a single cobalt ion in a CdTe/ZnTe QD and a single manganese ion in a CdSe/ZnSe QD. The spin states of the dopant ions are mapped onto the QD optical transitions recorded in a magneto-photoluminescence measurement.
We employ photoluminescence decay measurements to demonstrate that contrary to the case of systems with many magnetic ions\cite{Oka_JL99,Tang_JCG03,Lee_PRB05,Beaulac_NL08,Beaulac_NL08XPL,Zhong_Sp09,Bussian_NM09}, the exciton emission quenching is negligible for single dopants. Through modulated, polarization resolved photoluminescence measurements we access the single spin relaxation and prove that all-optical control of a single magnetic moment is feasible in the studied systems. Moreover, we show that spin properties of magnetic QDs can be designed by an independent choice of the magnetic ion and the QD material. We discuss the role of the electronic configuration of $d$-shell, spin-orbit and hyperfine interactions and, finally, we indicate the most promising design of future QD-based solotronic systems. \\

%#################################################### PL spectra

\begin{figure}[p]
\includegraphics[width=0.5\linewidth]{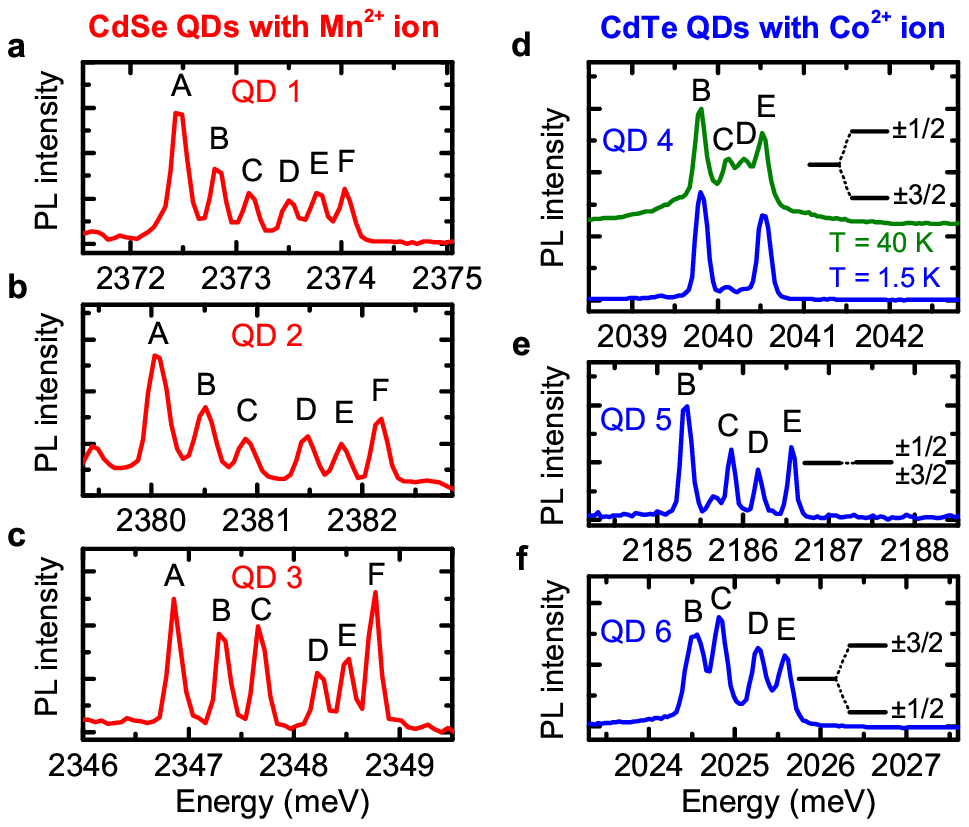}
\centering \caption{\textbf{PL spectra of CdSe QDs with single Mn$^{2+}$ ions (\textbf{a}, \textbf{b}, \textbf{c}) and CdTe QDs with single Co$^{2+}$ ions (\textbf{d}, \textbf{e}, \textbf{f}) at zero magnetic field and $T = 1.5$~K.}
The exciton splitting results from the $s$,$p$-$d$ exchange interaction with a magnetic ion.
For QD with Mn$^{2+}$ ion ($S=5/2$) we observe six components which in $\sigma^{\pm}$ circular polarization correspond to Mn$^{2+}$ spin projections $\mp5/2$ (A), $\mp3/2$ (B), $\mp1/2$ (C), $\pm1/2$ (D), $\pm3/2$ (E), $\pm5/2$ (F) while for QD with Co$^{2+}$ ion ($S=3/2$) there are four components related to Co$^{2+}$ spin projections $\mp 3/2$ (B), $\mp 1/2$ (C), $\pm 1/2$ (D), $\pm 3/2$ (E).
Intensities of various lines are related to a strain and resulting occupancy of magnetic ion states. For Mn$^{2+}$ effect of the strain is negligible (\textbf{a}, \textbf{b}, \textbf{c}) while for Co$^{2+}$ the impact is visible (\textbf{d}, \textbf{e}, \textbf{f}), e.g. cobalt shown in \textbf{d} has fundamental state with spin $\pm 3/2$, so observation of $\pm 1/2$ states requires an increse of the temperature. Upper curve in \textbf{d} was shifted to blue by 4.42 meV in order to compensate temperature shift.
 }
\label{fig:SoloZFS}
\end{figure}

%\section* {PL spectra of new QDs with single magnetic ions}
Samples with self-assembled QDs containing magnetic ions are grown by molecular beam epitaxy (see Methods) and studied by photoluminescence at low temperatures (down to 1.5 K). The use of a microscope objective enables the observation of exciton emission lines of individual QDs. Among sharp lines with a typical emission pattern of nonmagnetic QDs\cite{Kazimierczuk_PRB11} (majority) and broader lines related to QDs with many magnetic ions\cite{Tang_JCG03,Wojnar_PRB07,Klopotowski_PRB11}, it is possible to identify the emission multiplets characteristic for the individual QDs with exactly one magnetic ion, where the ion spin state is probed by confined excitonic complexes (neutral and charged excitons, biexciton, etc.).
In particular, the emission related to a bright state of neutral exciton (Fig. \ref{fig:SoloZFS}) consists of a set of lines split by the $s$,$p$-$d$ exchange interaction. The number of lines is determined by the possible magnetic ion spin projections on the growth axis being the exciton quantization axis. More specifically, for Mn$^{2+}$ with spin 5/2 there are 6 spin projections: $\pm 5/2$, $\pm 3/2$, and $\pm 1/2$, so we observe 6 lines for CdSe QD with the single Mn$^{2+}$ ion (Fig. \ref{fig:SoloZFS}a,b,c), analogously to a CdTe QD with the same ion\cite{Besombes_PRL04}. The Co$^{2+}$ ion spin is 3/2 and thus there are 4 spin projections: $\pm 3/2$ and $\pm 1/2$, resulting in 4 lines for CdTe QD with a single Co$^{2+}$ ion (Fig. \ref{fig:SoloZFS}d,e,f). However, the intensity of the lines related to Co$^{2+}$ spin projections $\pm 3/2$ (outer lines) can be significantly different from those related to the spin projections $\pm 1/2$ (inner lines). The Co$^{2+}$ ion orbital momentum is non-zero and Co$^{2+}$ incorporated to the crystal is very sensitive to a local anisotropy and the strain, which lead to the splitting of $\pm 3/2$ and $\pm 1/2$ states and a difference in their occupancy\cite{Macfarlane_JCP67,Koidl_PRB77,Pacuski_PRB06}. The QD shown in Fig. \ref{fig:SoloZFS}d, exhibits outer lines more intense than inner lines. This means that in this case the strain makes the state with spin $\pm 3/2$ the ground state. The difference between the spin state occupancies (and therefore the exciton line intensities) is more pronounced at low temperatures, as expected from Boltzmann statistics.

\begin{figure}[p!]
\includegraphics[width=1\linewidth]{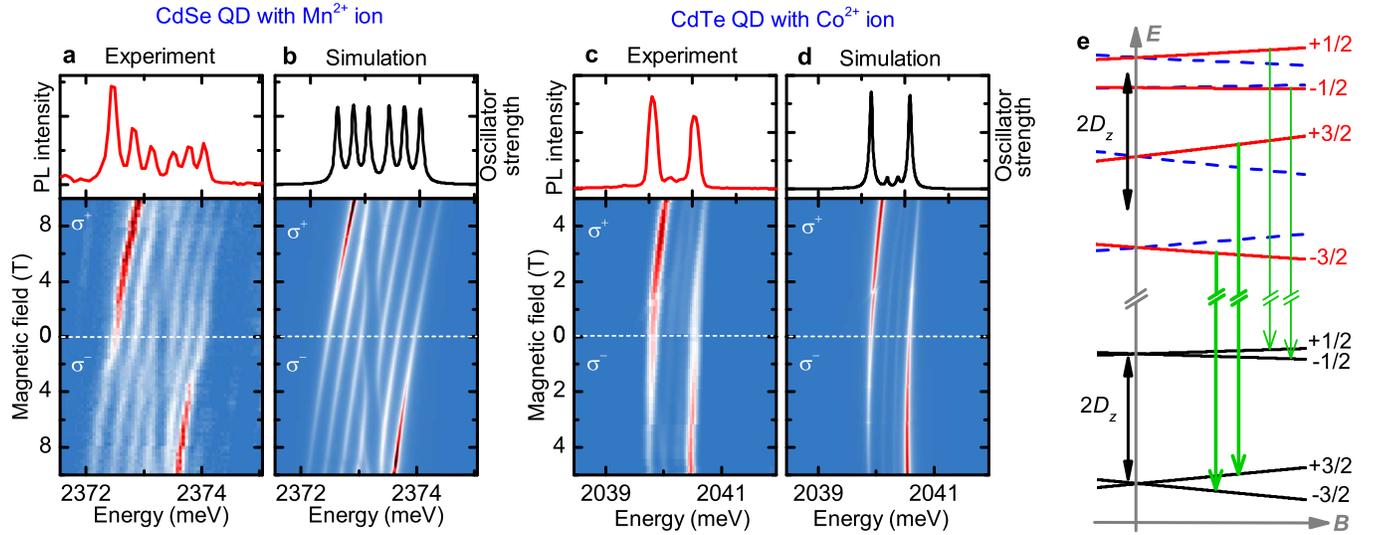}
\centering \caption{\textbf{Magnetooptical spectroscopy of QDs with single magnetic ions.} $T = 1.5$~K. PL of exciton in CdSe QD with Mn$^{2+}$ ion as a function of magnetic field in Faraday configuration (\textbf{a}) and a corresponding simulation (\textbf{b}). An analogous experiment (\textbf{c}) and simulation (\textbf{d}) for CdTe QD with Co$^{2+}$. More intense low-energy lines in the polarization $\sigma^{+}$ and high-energy lines in the polarization $\sigma^{-}$ at high magnetic field indicate alignment of ions spins along the external magnetic field direction. Parameters used for calculation of \textbf{b} and \textbf{d} are listed in Supplementary Table 1. Scheme of excitonic transitions (\textbf{e}) in  $\sigma^{+}$ polarization is shown for QD with Co$^{2+}$, for a relatively simple case, when strain induced Co$^{2+}$ anisotropy axis is parallel to the growth axis and resulting zero field spin splitting of Co$^{2+}$ states is equal $2D_z$, which is in microwave range, order of 1 meV. Spin projection of Co$^{2+}$ is indicated. }
\label{fig:SoloMO}
\end{figure}

The identification of the excitonic lines is confirmed by the analysis of photoluminescence spectra measured as a function of the magnetic field (Fig.~\ref{fig:SoloMO}). Zeeman shifts of bright and dark excitonic transitions for a CdSe QD with Mn$^{2+}$ (Fig.~\ref{fig:SoloMO}a) can be well described (Fig.~\ref{fig:SoloMO}b) using the model proposed for a CdTe QD with Mn$^{2+}$ [\onlinecite{Besombes_PRL04}]. In order to account for all the observed features of the PL spectrum from a CdTe QD with a single Co$^{2+}$ (Fig.~\ref{fig:SoloMO}c,d), we extend the model by introducing a strain vector that induces a zero field splitting of the Co$^{2+}$ spin states (see Methods). Fig.~\ref{fig:SoloMO}e shows the scheme of excitonic optical transitions for a relatively simple case, when the strain induced Co$^{2+}$ anisotropy axis is parallel to the growth axis. In plane anisotropy of Co$^{2+}$ would induce an additional excitonic mixing, analogous to the case of a neutral Mn center ($d^5+h$) in a InAs/GaAs QD\cite{Krebs_PRB09}. Typical photoluminescence spectra of various QDs with single magnetic ions are presented in Supplementary Figs. 1 and 2.\\

%########################################### Recombination

%\section* {Recombination channels of excitons}

Quenching of exciton photoluminescence by a recombination channel introduced by a magnetic dopant is an important obstacle in studying dilute magnetic semiconductors (DMS) with an energy gap larger than the magnetic ions internal transition energies\cite{Beaulac_NL08XPL,Zhong_Sp09,Yamamoto_JAP10}. This is the case of DMS with magnetic ions other than Mn (e.g. V, Cr, Fe, Co, Ni, Cu)\cite{Weakliem_JChP62,Baranowski_PR67,Koidl_PRB77}. Only Mn$^{2+}$ ions have relatively large intraionic transition energies (about 2.2 eV) allowing for efficient PL studies\cite{Cibert_book08,Goryca_PRL09QWs} of excitons in Cd$_{1-x}$Mn$_{x}$Te [energy gap E$_g$(CdTe) = 1.6 eV]. As a consequence, the first approaches to studies of dots with single magnetic ions were limited to manganese and QDs with low emission energy: CdTe/ZnTe\cite{Besombes_PRL04} and InAs/GaAs\cite{Kudelski_PRL07}. A relatively small energy gap of CdSe [E$_g$(CdSe) = 1.7 eV] results in  energy transfer from the magnetic ions to excitons\cite{Beaulac_NL08} and enables PL studies of bulk Cd$_{1-x}$Mn$_{x}$Se\cite{Giriat_PSSA80}. In the case of nanostructures, quantum confinement increases the exciton energy. As a consequence, the direction of the energy transfer is reversed, leading to PL quenching\cite{Beaulac_NL08,Beaulac_NL08XPL,Bussian_NM09,Tang_JCG03,Lee_PRB05} and a significant shortening of the exciton lifetime\cite{Oka_JL99,Yamamoto_JAP10}.

\begin{figure}[p]
\includegraphics[width=1\linewidth,clip]{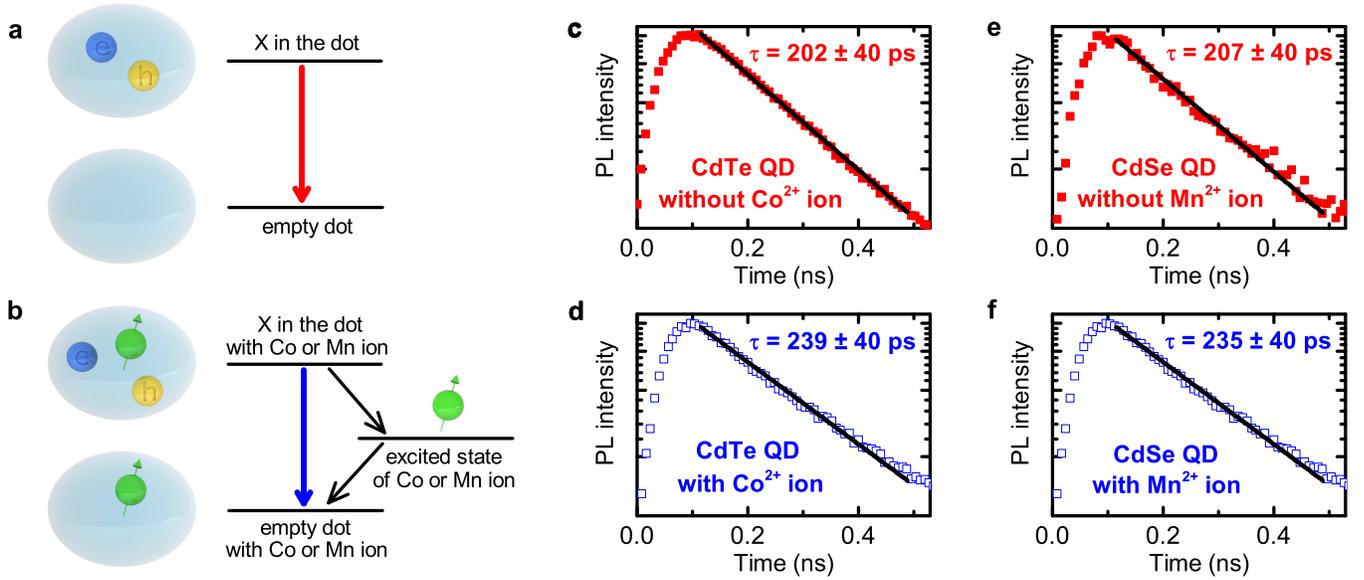}
\centering \caption{\textbf{Optical study of a QD confined exciton recombination} at $T=1.6$~K, $B=0$. Exciton recombination channels: \textbf{a} radiative in a nonmagnetic QD, \textbf{b} radiative and nonradiative in a QD with magnetic ion, which exhibit intra-ionic transitions at energies lower than exciton energy. Exciton PL decay measurement for CdTe QD (\textbf{c}), CdTe QD with a single Co$^{2+}$ (\textbf{d}), CdSe QD (\textbf{e}), and CdSe QD with a single Mn$^{2+}$ (\textbf{f}). We do not observe impact of single dopants on exciton decay, what means that the exciton decays mainly through the faster, radiative channel. Temporal resolution of the setup is 40 ps.
%\\\\\\\\\\\\\\\\\\\\
}
\label{fig:lifetime}
\end{figure}

However, our measurements of the PL decay performed on CdTe/ZnTe sample reveal the same lifetime of about 220$\pm40$ ps for a QD with and without a single Co$^{2+}$ ion (see Fig. \ref{fig:lifetime}c,d). Also for selenide QDs exciton lifetime is measured as around 220$\pm40$ ps for a dot with and without a single Mn$^{2+}$ (Fig. \ref{fig:lifetime}e,f). Therefore, we do not observe shortening of the exciton lifetime induced by single magnetic ions. Moreover, PL intensity of individual QDs with single magnetic ions is comparable to the one of nonmagnetic QDs. The above findings indicate that the quenching of excitonic emission is not efficient when single dopants are introduced to QDs. We interpret this as a result of a discrete density of states of zero dimensional systems with exactly one magnetic ion. In this case the exciton energy can not be efficiently transferred neither to phonons nor to the magnetic ion due to an energy mismatch. As a result the radiative channel of exciton recombination remains the most efficient one. This differs from the bulk DMS case, where the exciton is typically coupled to an ensemble of magnetic ions, able to absorb energy in a wide range - e.g., by a collective change of spin configuration. Above results not only show that PL studies of DMS can be significantly extended by using zero dimensional structures, but they imply also a high fidelity optical readout of a single dopant quantum states when it is embedded in a QD. \\

%########################################### Spin relaxation

%\section* {Single spin relaxation}

\begin{figure}
\includegraphics[width=0.8\linewidth]{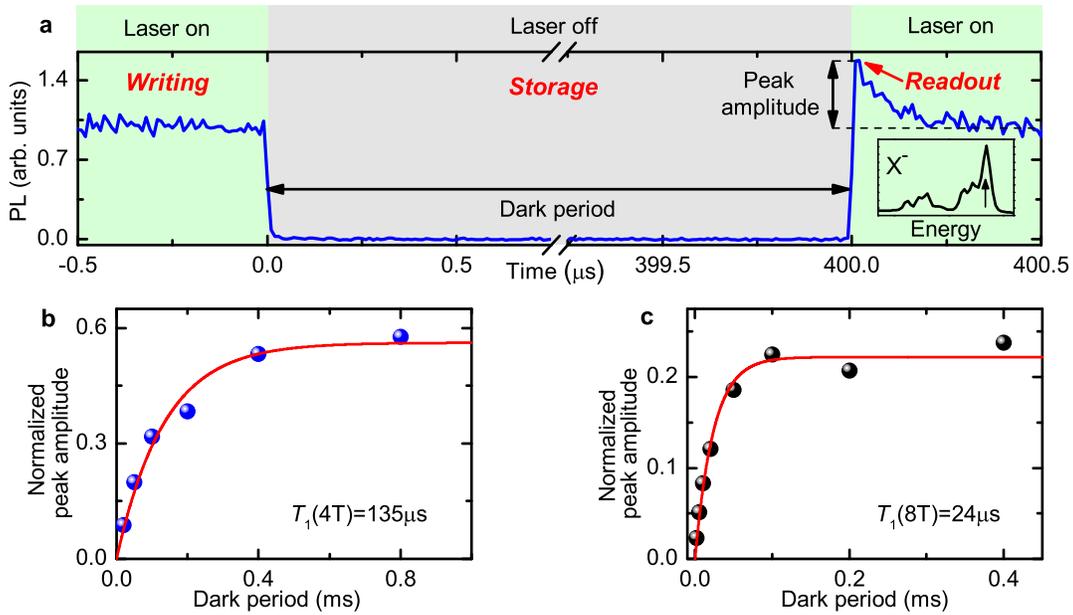}
\centering
\caption{\textbf{The measurement of the relaxation dynamics of a single Mn$^{2+}$ ion spin in CdSe/ZnSe QD at $T=1.6$~K.} \textbf{a} Temporal profile of the PL intensity of the highest energy X$^-$ emission line (indicated at inset) in magnetic field $B=4$ T (detection in $\sigma^-$ polarization). The nonresonant (405 nm) continuous-wave excitation is subsequently turned on and off for controlled periods of time. The amplitude of the PL intensity peak observed immediately after turning the excitation on indicates the loss of information stored on the Mn$^{2+}$ spin. \hspace{0.1cm} \textbf{b, c} The amplitude of PL intensity peak measured just after turning on the excitation versus length of the dark period for  external magnetic field of 4 T and 8 T, respectively. The fitted exponential curves yield the storage times $T_1$ of information on the Mn$^{2+}$ ion spin.}
\label{fig:spin_relaxation}
\end{figure}

In order to investigate spin relaxation dynamics of the Mn$^{2+}$ ion embedded in a CdSe/ZnSe QD, we measure time-resolved photoluminescence in a magnetic field, with an on/off modulated non-resonant laser excitation (Fig. \ref{fig:spin_relaxation}). When the laser is switched on, the optically created excitons injected to a QD depolarize the spin of the embedded magnetic ion (typically within several hundreds of nanoseconds). Then the laser is switched off for a dark period, during which the Mn$^{2+}$ spin approaches the state of alignment to the magnetic field direction. Finally, the laser is switched on again to perform the readout of the spin state by measuring the temporal increase of the PL amplitude of a high energy line. This line corresponds to the state of Mn$^{2+}$ ion with the spin oriented parallel to the magnetic field. Figures \ref{fig:spin_relaxation}b,c show the dependence of the PL amplitude on the length of the dark period for two different values of magnetic field. The exponential fits allow us to determine the Mn$^{2+}$ ion spin relaxation times in CdSe QD at helium temperature, in a magnetic field of 4~T and 8~T to be equal to 135$~\mu$s and 24$~\mu$s, respectively. It is over an \emph{order of magnitude longer} than the relaxation time of the Mn$^{2+}$ ion in CdTe QD, that is 5 $\mu$s for $B = 4$~T  measured by Goryca et al.\cite{Goryca2013} with the technique reported in Ref.~\onlinecite{Goryca_PRL09QDs}. Such results demonstrate that the modification of the QD material can significantly increase the storage time of information written on Mn$^{2+}$ spin, which can noticeably improve the features of quantum memories based on QDs with single magnetic ions. On the other hand, embedding an ion with a different electronic configuration may lead to an acceleration of the spin dynamics. In particular, similar measurements as described in Fig. \ref{fig:spin_relaxation} performed on CdTe QD with a single Co$^{2+}$ ion yields a relaxation time about 2 $\mu$s in a magnetic field of 3~T, shorter than for Mn$^{2+}$ ion in CdTe QD. \\

%########################################### Discussion

%\section* {Material systems for QDs based solotronics}

The results presented above highlight the opportunity of extension of QDs based solotronics research to yet unexplored combinations of semiconductor systems and magnetic ions. In this section we discuss the optimal systems from the point of view of practical implementation of QD based solotronic devices. Materials where ion-carrier exchange interaction has been experimentally observed are summarized in Table \ref{table}a. A similar Table \ref{table}b presents the most interesting combinations of QDs and magnetic ions. So far, only two systems were studied (single Mn in CdTe/ZnTe and InAs/GaAs QDs). With our work we introduce a new magnetic ion - Co$^{2+}$ and a new solotronic system - CdSe/ZnSe. Let's note first, that for applications in solotronics it is important to elongate the single spin relaxation time using magnetic ions without orbital momentum, so with half-filled $d$-shell ($d^5$). At non-zero magnetic fields, the Mn$^{2+}$ ions ($d^5$) in a bulk matrix  exhibit a relaxation time longer by two orders of magnitude than for other 3$d$ transition metals\cite{Blume_PR62}. This trend is confirmed in our work for  the ions in quantum dots, as Co$^{2+}$ ($d^7$ shell) exhibits a shorter relaxation time than Mn$^{2+}$. However, since increasing magnetic field enhances spin relaxation\cite{Blume_PR62} (see Fig. \ref{fig:spin_relaxation}), practical applications of solotronics devices will be realized at zero or low magnetic fields. At zero field the nuclear spin\cite{Goryca_PRL09QWs,Goryca_PRL09QDs,Gall_PRL09,Cao_PRB11,Gall_PRB12} plays the most important role and we predict, that other $d^5$ ion, Fe$^{3+}$ with zero nuclear spin should more stable than Mn$^{2+}$ with 5/2 nuclear spin. With Fe$^{3+}$ and QD build from isotopically purified group II and VI elements (which are attainable) in QDs, it will be possible to have a whole system free of nuclear spin and to obtain an extremely long spin relaxation time, as it was shown for N-V center in isotopically engineered diamond\cite{Balasubramanian_N09}.

The rate of spin relaxation of isolated ions increases with a strength of spin-orbit interaction\cite{Blume_PR62,Cibert_book08,Cao_PRB11}, which depends on the magnetic ion and a host material\cite{Wang_CJP10}. Therefore it is reasonable to use QDs based on semiconductors with light anions: sulfides, oxides, and nitrides, as they exhibit a weak spin orbit interaction. This trend\cite{Lambe_PR60} is confirmed by our results on QDs, as single Mn$^{2+}$ has a longer relaxation time when it is embedded into a selenide (Fig. \ref{fig:spin_relaxation}) than into a telluride\cite{Goryca_PRL09QDs,Goryca2013} (heavier) system. For ensembles of isolated Fe$^{3+}$ and Mn$^{2+}$ in oxides and sulfides long spin relaxation times up to 0.4 s [\onlinecite{Deville_JPF,Solomon_PR66}] and spin coherence time of 0.9 $\mu$s [\onlinecite{Ochsenbein_NN10}] were demonstrated at helium temperatures.

A manipulation of the stored spin counts equally for applications as the long spin relaxation time. The manipulation through ion-carrier exchange interaction\cite{Goryca_PRL09QDs} or a microwave radiation\cite{Ochsenbein_NN10} should be feasible for all the systems presented in Table \ref{table}b. Another, not exploited so far, possibility of the single magnetic ion spin manipulation in QD is related to intraionic transitions, which are similar to transitions exploited for defect centers\cite{Weber_PNAS10}. The intra-ionic transitions could be useful for reading, manipulation and writing of the magnetic ion spin. The spin readout from sharp intra-ionic transitions of Co$^{2+}$ ions in Zn$_{1-x}$Co$_x$O layers has been demonstrated in Ref. \onlinecite{Pacuski_PRB06}.

It is worth to note that the strain gives possibility of inducing a temporal evolution of spin, tuning zero field splitting, and reducing the spin degeneracy of the ground state. Thus, it can be profitable for manipulation of ions spins. The effect of the strain on Co$^{2+}$ spin states is shown in Fig. \ref{fig:SoloMO}e. In order to use the strain or crystal field effects, it is desired to control anisotropy axis, e.g. by using a wurtzite structure compounds, where the c-axis is expected to define the quantization axis of a zero field spin splitting. If one wants, in turn, to get rid of a strain induced complexity, lattice matched materials, like GaAs and AlAs for growth of magnetic QDs by a droplet epitaxy, are recommended. \\

In conclusion, optical properties of single Co$^{2+}$ ion in a CdTe QD and single Mn$^{2+}$ ion in a CdSe QD are presented for the first time. The QD emission decay time is found to be equal in the case of magnetic ion doped and undoped QDs. This indicates that quenching of the QD emission is negligible in zero-dimensional systems with single magnetic ions. A single manganese in the CdSe QD exhibits the longest spin relaxation time among the single magnetic ion-QD systems optically investigated so far. As such, it is clearly a good candidate for the single spin memory. However, the most promising systems for single spin memories have still not been tested experimentally: systems with a weak spin-orbit interaction, a wide energy gap, and with $d^5$ magnetic ions, thus self-assembled QDs based on oxides, sulfides or nitrides with single Mn$^{2+}$ or Fe$^{3+}$. Further experimental and theoretical studies are desired in order to indicate the optimal QD system for manipulation of single magnetic ion spin.\\
%########################################### Methods

\newpage
%\section*{Methods}
\textbf{Methods}\\

\textbf{Growth of samples with self-assembled QDs containing single magnetic ions.}
Samples are grown on GaAs (100)-oriented substrates using molecular beam epitaxy. After about 1 $\mu$m thick buffer layer of ZnTe (or ZnSe), about 3 monolayers of CdTe (or CdSe) were grown using atomic layer epitaxy by alternate cycles of Cd and Te (or Se). In the middle of CdTe (or CdSe) thin layer, a very small (previously calibrated by giant Zeeman splitting measurements of layers such as Zn$_{0.997}$Co$_{0.003}$Te) amount of magnetic ions: Co or Mn are introduced during 3 s of deposition. Next, the sample is cooled down for 1 h in Te (or Se) molecular flux, so at the end sample is covered with amorphous Te (or Se) which helps the QDs to form\cite{Tinjod_APL03,Wojnar_PRB07}. Subsequently the sample is heated up to the growth temperature, to sublimate the amorphous Te (or Se). The QDs are covered with a 100 nm ZnTe (or ZnSe) cap. No mesas or masks are needed to limit the number of observed QDs. The emission from single QDs at temperature of 2 K is collected with a microscope objective assuring a resolution of 0.5 $\mu$m.\\

\textbf{Modeling of magneto-photoluminescence.}
A quantitative description of measured QDs magneto-PL spectra is provided by a model of the neutral exciton inside a QD with a single magnetic dopant in magnetic field parallel to the growth axis. We consider initial and final states of photoluminescence transition (Fig. \ref{fig:SoloMO}e). For the initial state, the model takes into account the energy of an exciton in a nonmagnetic QD (isotropic and anisotropic electron-hole exchange interactions, heavy-light hole mixing, Zeeman effect and diamagnetic shift), energy of ion-electron and ion-hole exchange interactions ($s$,$p$-$d$ interactions), and finally the magnetic ion energy determined by the Zeeman effect and the strain vector. For the final state, we consider only the magnetic ion energy. Oscillator strength of calculated  optical transitions is multiplied by Boltzmann distribution of magnetic ion state occupancy and the resulting intensity is plot in Fig. \ref{fig:SoloMO}b,d. Its similarity to measured spectra can be considered as a positive test of the model used, Fig. \ref{fig:SoloMO}a,c. A more precise description of the Hamiltonian is given in Supplementary Information.\\

%############################################# Acknowledgements

\textbf{Acknowledgements}\\
We acknowledge helpful discussions with T. Dietl, T. Kazimierczuk {\L}. K\l{}opotowski, and {\L}. Cywi\'{n}ski. This work was partially supported by the Polish National Science Centre under decisions DEC-2011/01/B/ST3/02406, DEC-2011/02/A/ST3/00131, DEC-2012/05/N/ST3/03209, by the NCBiR project LIDER, and by Polish Ministry of Science and Higher Education as research grants ''Diamentowy Grant''. Project was carried out with the use of CePT, CeZaMat and NLTK infrastructures financed by the European Union - the European Regional Development Fund within the Operational Programme "Innovative economy" for 2007-2013. The work was also supported by the EC through the FunDMS Advanced Grant of the European Research Council (FP7 "Ideas").\\

\textbf{Author contributions}\\
J.K., M.P., K.G., J.-G.R., E.J., W.P. grew and characterized samples, J.K., T.S., M.K., M.G., A.G., P.K., W.P. performed magnetooptical experiments, data analysis, and modeling, T.S., A.B., M.G. performed single spin relaxation measurements, T.S., J.K., J.S., M.N., A.G., W.P. prepared the manuscript in consultation with all authors.\\

\textbf{Author information}\\
Correspondence and requests for materials or samples should be addressed to Wojciech.Pacuski@fuw.edu.pl \\

\textbf{Competing financial interests}\\
The authors declare no competing financial interests

%############################################# Figures

\newpage
%\bibliographystyle{naturemag}
%\bibliography{bibsolotronics}

\begin{table}[p]
\includegraphics[width=1\linewidth]{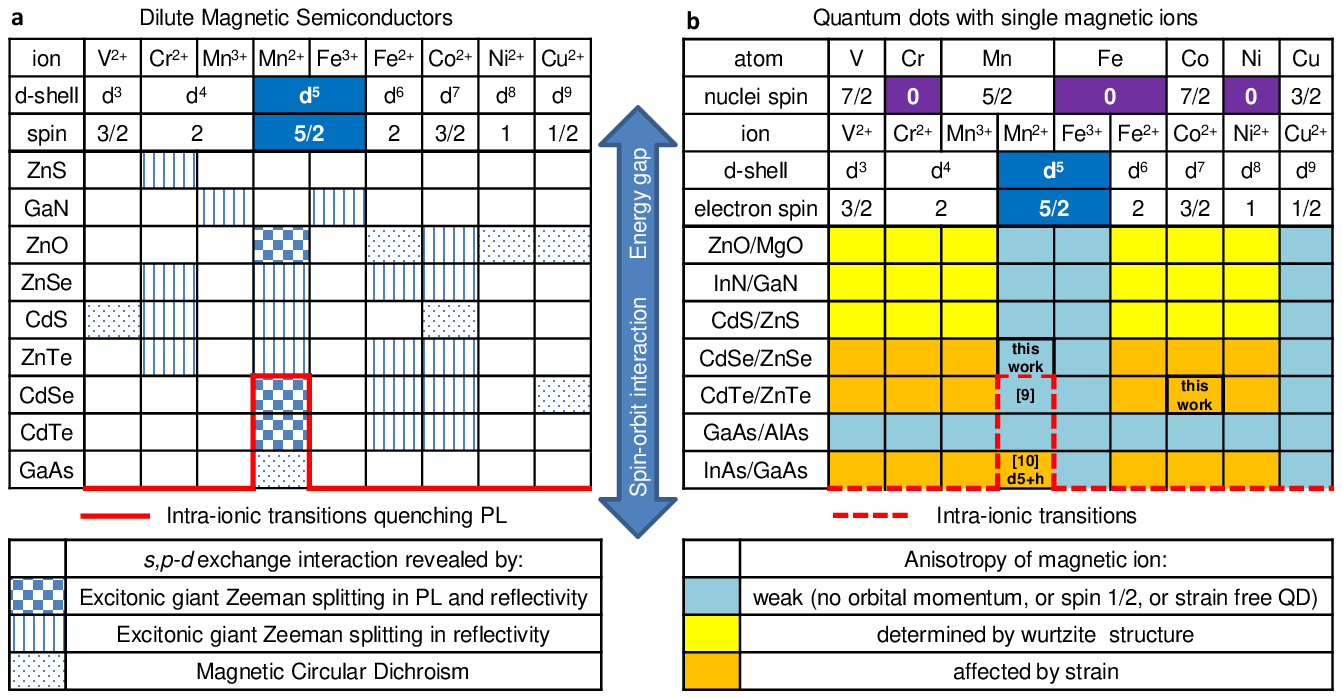}
\centering \caption{\textbf{Tables summarizing semiconductor systems where interaction of excitons and magnetic ion has been confirmed using optical spectroscopy.} \textbf{a} Dilute Magnetic Semiconductors (DMS). \textbf{b} Quantum dots with single magnetic ions. Vertical order is given by increasing energy gap which is anti-correlated with a strength of spin-orbit interaction of a host semiconductor (from weak to strong spin-orbit coupling are systems with anions O, N, S, Se, As, Te). Horizontal order is given by $d$-shell occupancy of magnetic dopants. The area above red line corresponds to materials exhibiting intra-ionic transitions which quench PL in DMS, but affect only weakly PL in QDs with single magnetic ions (as this work shows). Detailed references list for DMS is given in Supplementary Table 2. QDs with single magnetic ions were reported so far only for two systems CdTe/ZnTe QDs with Mn$^{2+}$ [\onlinecite{Besombes_PRL04}] and InAs/GaAs QDs with complex of Mn$^{2+}$ and bound hole [\onlinecite{Kudelski_PRL07}]. This work reports on two more systems and indicates that majority of displayed combinations should be useful for optical spectroscopy and for applications.}
\label{table}
\end{table}

\end{document}

%\preprint{Warsaw, November 1, 2012}
%\date{\today}
%################################################################# TITLE

\title{Designing quantum dots for solotronics - supplementary information}

%################################################################# AUTHORS

\author{J.~Kobak}
\author{T.~Smole\'nski}
\author{M.~Goryca}
\author{M.~Papaj}
\author{K.~Gietka}
\author{A.~Bogucki}
\author{M.~Koperski}
\author{J.-G.~Rousset}
\author{J.~Suf\mbox{}fczy\'{n}ski}
\author{E.~Janik}
\author{M.~Nawrocki}
\author{A.~Golnik}
\author{P.~Kossacki}
\author{W.~Pacuski}
%\email{Wojciech.Pacuski@fuw.edu.pl}

\affiliation{Institute of Experimental Physics, Faculty of Physics, University of Warsaw, Ho\.za 69, 00-681 Warsaw, Poland}

\maketitle

%################################################################# ABSTRACT

%\begin{abstract}
This supplementary information contains: (i) example zero-field PL spectra of various QDs with single magnetic ions, (ii) details of the model used for simulation of magneto-photoluminescence presented in the main text, and (iii) complementary information to the table of dilute magnetic semiconductors presented in the main text.\\
%\end{abstract}

%#################################################################
\noindent
\textbf{Zero field PL spectra of various QDs with single magnetic ions}\\

In supplementary Figs. S1  and S2 are presented examples of photoluminescence spectra of individual CdSe and CdTe QDs with single magnetic ions (Mn$^{2+}$ and Co$^{2+}$) and without magnetic ions. At low excitation power the line of neutral exciton (X) and one or two lines of charged excitons (X$^{-}$ and X$^{+}$) are obvserved. At higher excitation power, it is usually possible to resolve biexciton line (XX) and lines related to higher excitonic complexes (not marked in Figs. S1 and S2). The order in emission energy (from higher to lower energies) is: the exciton line, the X$^{+}$, next the X$^{-}$, and finally the XX. Such order is kept for all the observed QDs, regardless of the QD material. Since selenides tend to be n-type materials we do not observe, however, positively charged exciton line for CdSe/ZnSe QDs. Exciton - biexciton separation energy is about 11 meV for CdTe QDs and about 24 meV for CdSe QDs.

Identification of excitonic emission lines for nonmagnetic CdTe and CdSe QDs was extensively discussed in literature\cite{Kulakovskii_PRL99,Patton_PRB03,Suffczynski_PRB06,Kazimierczuk_PRB11,Kazimierczuk_PRB13}. In order to confirm our identification, in addition to the comparison  of the relative positions of the excitonic transtions to the literature ones,  dependence of the line position on the detected linear polarization angle, line intensity dependence on the excitation power, and evolution of the excitonic lines in magnetic field was analyzed. QDs with single magnetic ions exhibit lines split by $s$,$p$-$d$ exchange interactions. The neutral exciton and biexciton lines typically exhibit 6 components for Mn$^{2+}$ (spin 5/2) and 4 components for Co$^{2+}$ (spin 3/2) as marked in the figures. Intensity of various components of exciton in CdTe QD with Co$^{2+}$ depends on strain and temperature. In some cases mixing with dark-excitons\cite{Besombes_PRL04,Goryca_PRB10} leads to observation of more lines for neutral exciton, as shown in the Supplementary Figs. S1c and S2f.

%\clearpage

\begin{figure*}[h!]
\includegraphics[width=0.643\linewidth,clip]{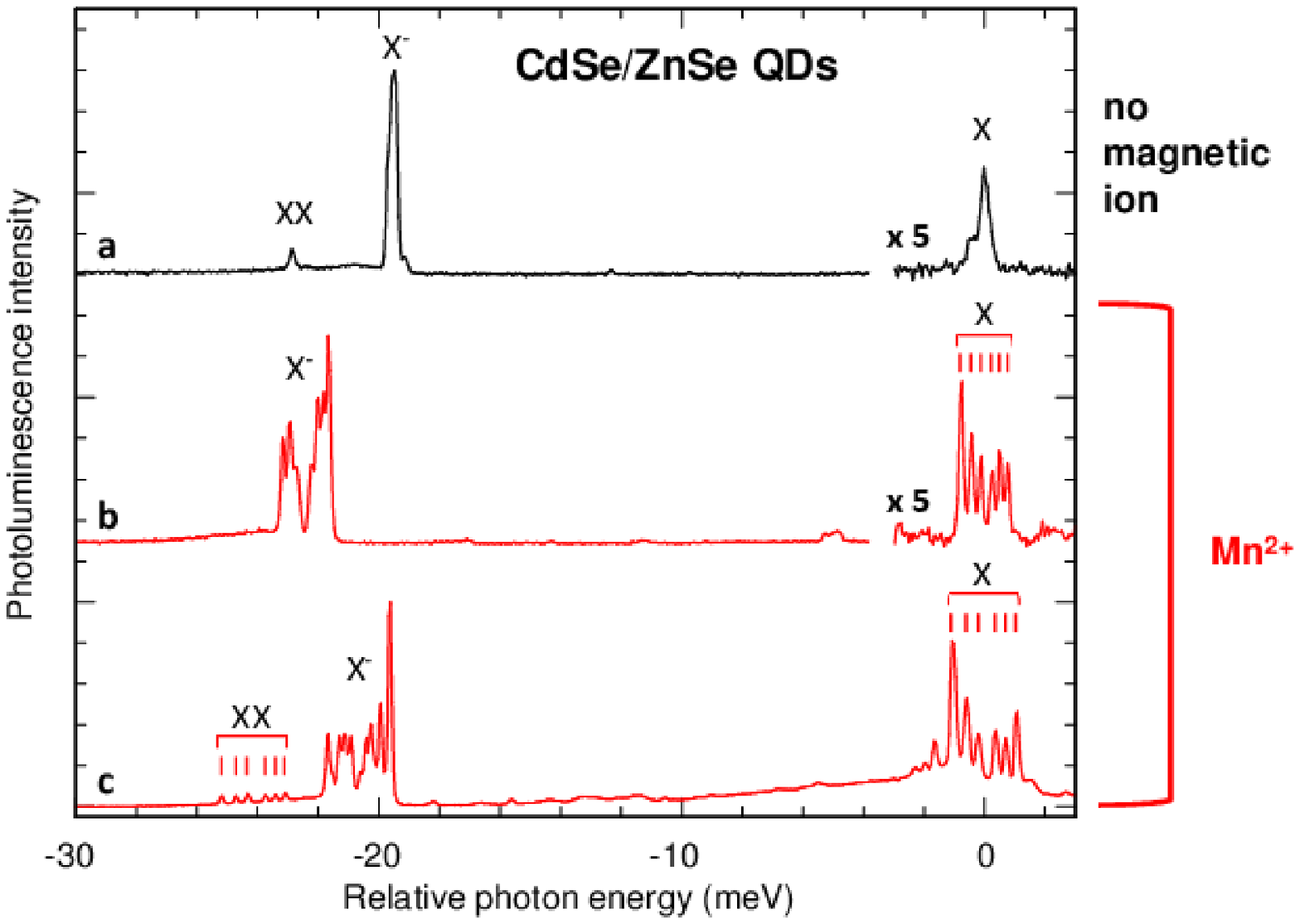}
%\centering
%\\[0.3cm]
\label{fig:soloCdSeQdsXrescaled}
\end{figure*}
\noindent
\textbf{Supplementary Fig. 1: PL spectra of individual CdSe QDs at $T=1.5$~K}. Lines related to neutral exciton (X), charged exciton (X$^{-}$) and biexciton (XX) are marked.\hspace{0.05cm}  \textbf{a} QD without magnetic ion. \hspace{0.05cm} \textbf{b,c} QDs with single Mn$^{2+}$. Principal six lines of biexciton and neutral bright exciton originates from $s$,$p$-$d$ exchange interaction of exciton with Mn$^{2+}$ ion. Additional weak lines below X in (\textbf{c}) are related to dark-exciton states. Energies of the neutral exciton emission are equal to 2318 meV (\textbf{a}), 2373 meV (\textbf{b}), and 2381 meV (\textbf{c}).

\clearpage

\begin{figure*}[h!]
\includegraphics[width=0.643\linewidth]{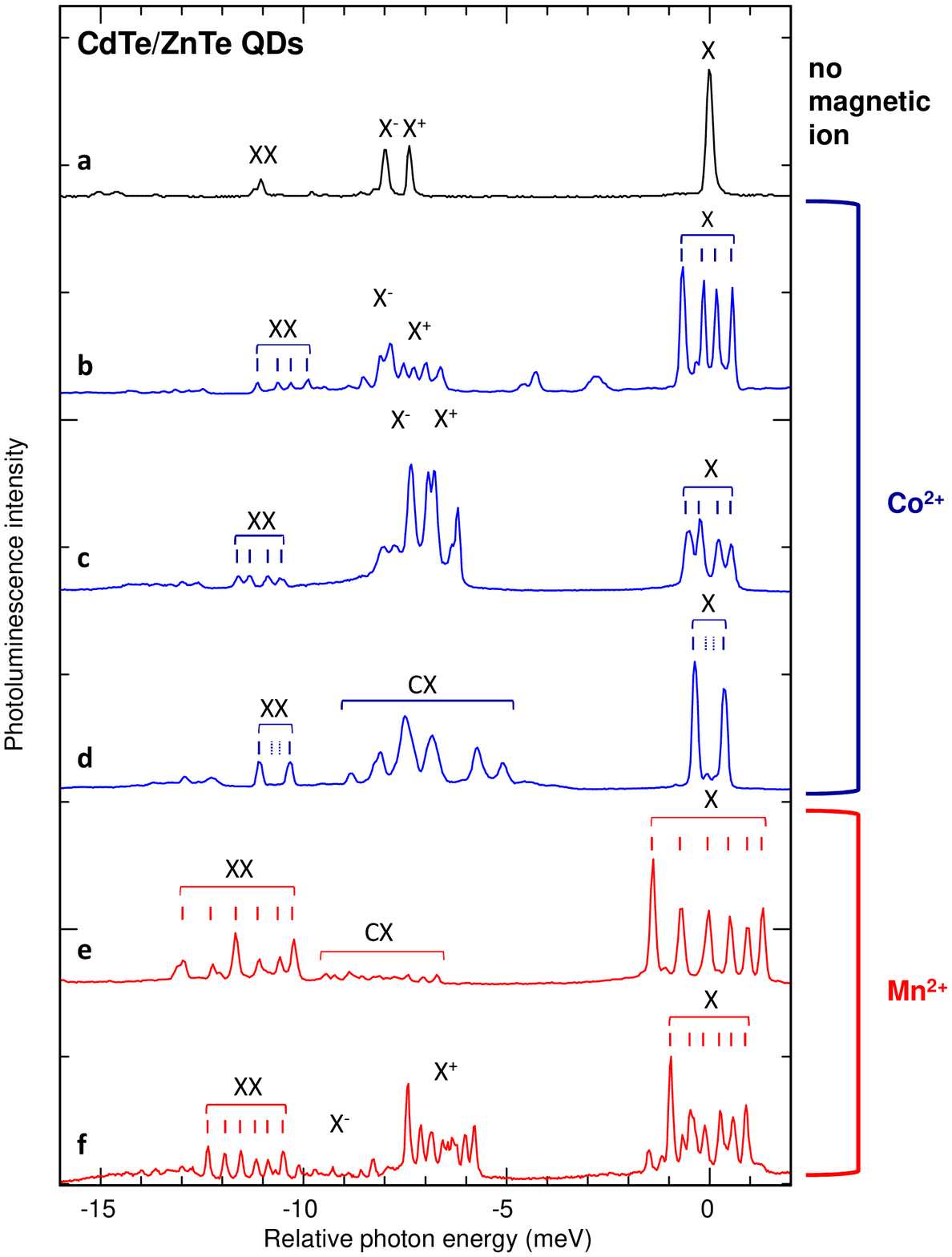}
%\centering
%\\[0.3cm]
\label{fig:SoloCdTeQDs}
\end{figure*}
\noindent
\textbf{Supplementary Fig. 2: PL spectra of individual CdTe QDs, at $T=1.5$~K}. Lines related to neutral exciton (X), charged excitons (X$^{-}$ and X$^{+}$), overlapping charged excitons (CX) and biexciton (XX) are marked.
\hspace{0.05cm}  \textbf{a} QD without magnetic ion. \hspace{0.05cm} \textbf{b,c,d} QDs with single Co$^{2+}$. \hspace{0.05cm} \textbf{e,f} QDs with single Mn$^{2+}$. Due to exchange interaction between exciton and magnetic ion neutral exciton and biexciton states split and we observe four related emission lines in case of QDs with a single Co$^{2+}$ (\textbf{b,c,d}) or six emission lines in case of QDs with a single Mn$^{2+}$ (\textbf{e,f}). Additional weak lines below X in (\textbf{f}) are related to a strong mixing of dark-exciton and bright-exciton states. In spectra (\textbf{b,c,d}), intensity of various components of exciton in CdTe QD with Co$^{2+}$ depends on strain induced zero-field splitting of Co$^{2+}$ spin states $\pm1/2$ and $\pm3/2$. Energies of the neutral exciton emission are equal to 2185 meV (\textbf{a}), 2186 meV (\textbf{b}), 2025 meV (\textbf{c}), 2040~meV (\textbf{d}), 2214 meV (\textbf{e}), and 1961 meV (\textbf{f}).

\clearpage
%#################################################################
\newpage
\newpage
\newpage
\noindent
\textbf{Details of magneto-photoluminescence modelling }\\

Our description of the neutral exciton inside a QD with a single magnetic dopant,
in magnetic field ($B$) parallel to the growth axis $z$, is given by the following
Hamiltonian :
\begin{eqnarray*}
\mathcal{H}_{X,M}(B)&=&\mu_BB\left(g_eS^z_e+g_hS^z_h\right)+I_e\vec{S}_e\vec{J}+I_h\vec{S}_h\vec{J}-\frac{2\delta_0}{3}S^z_eS^z_h
+\frac{2\delta_1}{3}\left(S^y_e(S^y_h)^3-S^x_e(S^x_h)^3\right)+\\
&+&\Delta_{lh-hh}\left((S^x_h)^2-(S^y_h)^2\right)+\rho\left((S^x_h)^2+(S^y_h)^2\right)+\gamma B^2 + \mathcal{H}_M,\\[0.3cm]
\mathcal{H}_M&=&\mu_BBg_MJ^z +D_z(J^z)^2+D_x(J^x)^2+D_y(J^y)^2,
\end{eqnarray*}

where $\mathcal{H}_M$ is the Hamiltonian of the magnetic ion, $g_e$, $g_h$, $g_M$ are electron, hole and
magnetic ion $g$ factors, $\vec{S}_e$, $\vec{S}_h$ and $\vec{J}$ are electron, hole and
magnetic ion spin operators, $I_e$ and $I_h$ are the electron-ion and
hole-ion exchange constants\cite{Gaj_book10,Besombes_PRL04}, $\delta_0$ and $\delta_1$ are the energies
related to isotropic and anisotropic parts of the electron-hole exchange
interaction\cite{Bayer_PRB02}, $\Delta_{lh-hh}$ is the heavy-light hole splitting, $\rho$ represents
the strength of the valence-band mixing\cite{Leger_PRB07,Koudinov_PRB04} and $\gamma$ is an excitonic
diamagnetic shift constant. The parameters $D_x$, $D_y$, $D_z$ describe strain induced zero-field splitting and the anisotropy axis of Co$^{2+}$ ion spin states.  In the case of Mn$^{2+}$ ion, strain effects\cite{Qazzaz_SSC95} are much weaker and zero field splitting is neglected. Supplementary Table 1 lists example parameters of the Hamiltonian. This parameters were used for calculation of spectra of CdSe QD with single Mn$^{2+}$ and CdTe QD with single Co$^{2+}$ presented in Figs. 2b,d of the main text. Parameters $g_M$ are literature data\cite{Ham_PRL60, Title_PR63}, other parameters were determined from fit to data presented in Figs. 2 a,c.

Similar fits and determination of comparable excitonic parameters were already reported and discussed in literature devoted to CdTe/ZnTe QDs with single Mn\cite{Besombes_PRL04,Besombes_APPA05,Besombes_PRB05,Goryca_PRB10}, nonmagnetic QDs CdTe/ZnTe\cite{Kazimierczuk_PRB11,Smolenski_PRB12,Kazimierczuk_PRB13,Hewaparakrama_NT08}, and nonmagnetic QDs CdSe/ZnSe\cite{Akimov_PRB05,Hundt_PSSB}. New in this work is determination of parameter $D$ for single ion in QD. We note that for QD discussed in this work,  which is rather typical CdTe QD with Co$^{2+}$, $D_z= -1.4$~meV, so it is negative, and its absolute value is larger than $D$ values observed in wurtzite structure DMSs with Co$^{2+}$: 0.062 meV for CdSe\cite{Lewi91}, 0.084 meV for CdS\cite{Lewi91}, and 0.34 meV for ZnO\cite{Jedr04, Sati06, Pacuski_PRB06}.

Oscillator strength of optical transitions are calculated for initial state given by $\mathcal{H}_{X,M}$ and final state given by $\mathcal{H}_{M}$. Line intensity is calculated as a multiplication of oscillator strength and magnetic ion spin state occupancy given by Boltzmann distribution for an effective temperature T$_{eff}$. In this simple model relaxation of exciton-ion complex is neglected.\\

\begin{table}[h!] %-----------------------------------------------
%\centering
%\hline
\setlength{\tabcolsep}{6.5pt}
\begin{tabular}{lcccccccccccccccccc}
%\vspace{0.1cm}

Ion & QD & $I_e$ & $I_h$ & $\delta_0$ & $\delta_1$ & $\frac{\rho}{\Delta_{lh-hh}}$ & $\gamma$ & $g_e$ &  $g_h$ & $g_M$  & $D_x$ & $D_y$ & $D_z$ & $T_{eff}$\\[0.2cm]

 & & meV & meV & meV & meV &   & $\mu$eV/T$^2$ &   &    &    & meV & meV & meV & K\\[0.2cm]

\hline
\\[-0.2cm]

Mn$^{2+}$ & CdSe& -0.015 & 0.195 & 1.0 &  0.23 & 0 & 0.5 & -0.4 &  0.35 & 2.0  & 0 & 0 & 0 & 30\\[0.2cm]

Co$^{2+}$ & CdTe& -0.06 & 0.16 & 0.8 &  0.01 & 0.5 & 3.2 & -0.35 &  0.2 & 2.3  & 0 & -0.3 & -1.4  & 10\\[0.2cm]

%\hline \\[0.3cm]

\end{tabular}

\label{tab:QDs}
\end{table}

\noindent
\textbf{Supplementary Table 1: Parameters of the model used in calculation of QDs spectra presented in Figs. 2b,d of the main text.}\\

\newpage
\noindent
\textbf{Reported magneto-optical study of Dilute Magnetic Semiconductors}\\

A characteristic property of DMSs is an interaction of magnetic ions and band carriers, so called $s$,$p$-$d$ exchange interaction, which result in giant Zeeman splitting of excitons and in enhancement of magnetic circular dichroism (MCD). Supplementary Table 2 summarizes magneto-optical study which allowed for determination of effective $s$,$p$-$d$ exchange integrals in bulk semiconductors and MCD study evidencing $s$,$p$-$d$ exchange interaction in nanocrystals with magnetic ions. Significant number of studied DMS systems indicates space for engineering QDs with single magnetic ions. Motivation of this work resulted from our previous study of Co$^{2+}$ based DMSs\cite{Nawr91, Pacuski_PRB07, Papaj_APPA12}. Before our work, only Mn$^{2+}$ was used as a single magnetic ions in CdTe QDs\cite{Besombes_PRL04,Besombes_PRB05,Leger_PRL05,Maingault_APL06,Leger_PRL06,XJLi_PE08,Gall_PRL09,Gall_PRL12,Gall_PRB10,Goryca_PRB10,Goryca_PRL09QDs,Wojnar_PRB07,Fernandez_PRL07,Gall_PRB12,Gietka_APPA12} and InAs QDs\cite{Kudelski_PRL07,Krebs_PRB09,Baudin_PRL11}. This work presents two new systems, CdSe QDs with Mn$^{2+}$ and CdTe QDs with Co$^{2+}$, and indicates that majority of displayed combinations should be useful for optical spectroscopy as well as for practical applications.
\\

\begin{table}[h!] %-----------------------------------------------
%\centering
%\hline
\setlength{\tabcolsep}{8pt}
\begin{tabular}{lccccccccc}
%\vspace{0.1cm}

Ion & V$^{2+}$ & Cr$^{2+}$ & Mn$^{3+}$ &  Mn$^{2+}$ & Fe$^{3+}$ & Fe$^{2+}$ & Co$^{2+}$ & Ni$^{2+}$ & Cu$^{2+}$\\[0.2cm]
$d$-shell& $d^3$ & $d^4$ & $d^4$ & $d^5$ & $d^5$ & $d^6$ & $d^7$ & $d^8$ & $d^9$ \\[0.2cm]
\hline
\\[-0.2cm]
ZnO & & & &\onlinecite{Prze06,Pacuski_PRB11} & & \onlinecite{Ando04}&\onlinecite{Pacuski_PRB06} &\onlinecite{Ando_JAP01}$^*$ &\onlinecite{Ando_JAP01}$^*$ \\[0.1cm]
ZnS & &\onlinecite{Mac96}& & & & & & & \\[0.1cm]
ZnSe & & \onlinecite{Mac96}& &\onlinecite{Twar84znse} & & \onlinecite{Twar90}&\onlinecite{LiuPetrou90} & &\\[0.1cm]
ZnTe & & \onlinecite{Mac96}& &\onlinecite{Twar84znte} & & \onlinecite{Test00}&\onlinecite{Ziel96} & & \\[0.1cm]
CdS &\onlinecite{Mac_SST00} & \onlinecite{Herb98}& &\onlinecite{Nawr87} & & & \onlinecite{Radovanovic_02}$^*$\\[0.1cm]
CdSe & &  & &\onlinecite{Arci86} & &\onlinecite{Scal90}&\onlinecite{Nawr91}  & &\onlinecite{Pandey_NM12}$^*$\\[0.1cm]
CdTe & &  & &\onlinecite{Gaj79}& & \onlinecite{Test91} & \onlinecite{Ziel00,Alaw01}\\[0.1cm]
GaN & & &\onlinecite{Pacuski_PRB07,Suffczynski_PRB10}&&\onlinecite{Pacuski_PRL08}\\[0.1cm]
GaAs & &  & &\onlinecite{Szczytko_PRB99}\\[0.2cm]

\hline \\[0.3cm]

\end{tabular}

%\caption[Dilute Magnetic Semiconductors]
\label{tab:DMSs}

\end{table}
\noindent
\textbf{Supplementary Table 2: Summary of references related to the magneto-optical observation of the excitonic giant Zeeman splitting in bulk dilute magnetic semiconductors.} Additionally, references related to MCD study of nanocrystals with magnetic ions are shown with a star ($^*$).  The Table is an updated set of data from Ref. \onlinecite{Gaj_book10}.

%\bibliographystyle{naturemag}
%\bibliography{bibsolotronics}